\documentclass[prb,twocolumn,aps,showpacs]{revtex4}%
\usepackage{amssymb}
\usepackage{graphicx}
\usepackage{longtable}
\usepackage{amsmath}
\usepackage{amsfonts}%
\providecommand{\U}[1]{\protect\rule{.1in}{.1in}}
\begin{document}
\author{E. Pavarini }
\affiliation{INFM-Dipartimento di Fisica \textquotedblleft A.Volta\textquotedblright, }
\affiliation{Universit\`{a} di Pavia, I-27100 Pavia, Italy}
\affiliation{Institut f\"ur Festk\"orperforschung, Forschungzentrum J\"ulich, D-52425
J\"ulich (Germany)}
\author{I. I. Mazin}
\affiliation{Code 6390, Naval Research Laboratory, Washington, DC 20375}
\date{\today }

\title{First-principles study of spin-orbit effects and NMR in Sr$_{2}$RuO$_{4}$}

\begin{abstract}
{We present a first principles study of NMR and spin orbit effects in the unconventional
superconductor Sr$_{2}%
$RuO$_{4}$. We have calculated the uniform magnetic susceptibility,
which agrees rather well with the experiment in amplitude, 
but, as in an earlier model result
we found  the calculated hard axis  to be $z$, opposite to the experiment.
We have also calculated the Knight shifts and the NMR relaxation rates
for all atoms, and again found an overall good agreement, but important 
deviations from the experiment in same particular characteristic, such as
the Knight shift anisotropy. Our results suggest that correlations in Sr$_{2}%
$RuO$_{4}$ lead
to underestimations of the orbital effects in density-functional based
calculations. We also argue that the accepted ``experimental'' value
for the relative contribution of orbital polarization in susceptibility,
10-15\%, is also an underestimation.
We  discuss the puzzling invariance of the
the O and Ru Knight shift 
across the superconducting transition for all directions of 
the applied field.
We show that this fact
cannot be explained by accidental cancellations or spin-flip
scattering, as it happens in some elemental superconductors.
We also point out that large contribution of the dipole and orbital
hyperfine field into the Knight shifts in  Sr$_{2}%
$RuO$_{4}$, combined with the possibility of an orbital-dependent
superconductivity, calls for a revision of the standard theory of the
Knight shift in the superconducting state.
}

\end{abstract}
\maketitle

\section{Introduction}

The layered perovskite Sr$_{2}$RuO$_{4}$ is often considered to be one of the
rare cases of spin-triplet superconductors.\cite{Mac} (see, however, Ref.~\onlinecite{new}). 
While there is convincing evidence against a conventional
$s$-wave state\cite{AG} the evidence for a spin-triplet $p$-wave
superconductivity rests exclusively\cite{new} on the NMR experiments.
Specifically, the main (and, arguably, the only) argument in favor of a
(particular) $p$-wave state is the fact that the NMR Knight shift for a
magnetic field parallel to the RuO$_{2}$ layers does not decrease in the
superconducting state. Indeed, in a singlet superconductor the Cooper pairs
have no net spin and cannot produce any Knight shift, so for applied fields
smaller, in energy units, than the superconducting gap, as the number of
unpaired electrons at the Fermi level decreases so does the Knight shift. On
the contrary, among the $p-$wave states allowed in a tetragonal symmetry,
there is one where the spins of the pairs lie in the $ab$ plane and are
capable of screening an external in-plane field. Indeed, a simple
theory\cite{Ktheory} predicts that the Knight shift for such fields remains
constant across the superconducting transition. The latter appears to be well
established, being confirmed for the $^{17}$O shift \cite{Ishida98}, for the
$^{101}$Ru shift \cite{Ishida01}, and for the spin susceptibility as probed by
neutrons \cite{Duffy00}.

The superconducting state in question, sometimes called a chiral $p$-wave
state, is described by a vector order parameter $\mathbf{d\propto(}k_{x}\pm
ik_{y})\mathbf{\hat{z},}$ and is also compatible with the $\mu$SR relaxation
experiments\cite{uSR} which indicate appearance of spontaneous magnetic
moments below $T_{c}$ that is usually interpreted in terms of a
superconducting state with Cooper pairs having nonzero orbital moments. On the
other hand, it is not readily compatible with multiple indications of line
nodes of the order parameter\cite{Mac}, since such lines are not required by
symmetry in this chiral $p$-wave state and can appear only accidentally. While
such an accident is always possible, no superconductors are  known so far
where the order parameter would have nodes not required by symmetry, for an
obvious reason that such nodes strongly reduce the total pairing energy.

The confidence in the chiral $p$-wave state has been further shaken by the
fact that the Knight shift in Sr$_{2}$RuO$_{4}$ remains constant not only for
the in-plane fileds, but in fact for $any$ direction of the applied
field\cite{Murukawa04}! According to the existing theory, no superconducting
state allowed for tetragonal crystals can have such a property. The suggested
explanation was that the direction of \textbf{d} in a magnetic field of 0.02 T
changes, \cite{Murukawa04} creating a state with the symmetry \textbf{d=}%
$k_{z}{\mathbf{\hat{x}}}$ [a \textquotedblleft rotated\textquotedblright%
\ state \textbf{d=}$(k_{z}+ik_{y}){\mathbf{\hat{x}}} $ is not allowed in a
tetragonal symmetry\cite{UedaSigrist} and can appear only as a second phase
transition]. This explanation, however, is rather doubtful for the following
reasons: (i) such a state would have an additional node line and therefore
loose a good deal of pairing energy and (ii) although in this state the spins
of the pairs lie in the $yz$ plane, there is no $y-z$ symmetry (as opposed to
the $xy$ plane) and it is not $a$ $priori $ clear whether the magnetic
susceptibility of the Cooper pair will be the same as for the normal
electrons. Finally, (iii) spin-orbital part of the pairing interaction, which
keeps the spins in the $xy$ plane, despite $z$ being the easy magnetization
axis\cite{Maeno97}, is rather strong in this material, and the field of 0.02 T
(1.1 $\mu$eV or 0.013 K) seems to be way too small to overcome it.

These considerations have spawn several alternative suggestions for the pairing 
symmetry in Sr$_2$RuO$_4$, such as a chiral $d$-wave state\cite{new} or 
a mixture of nearly-degenerate planar $p$-wave states\cite{Dan}. 
On the other hand, examples are known where the magnetic
susceptibility and the Knight shift do not vanish at $T=0$ even in conventional
superconductors: mercury, \cite{Reif} tin \cite{Androes} or 
vanadium.\cite{Noer} Such cases are traditionally attributed to either
spin-flip scattering due to spin-orbit coupling on point
defects\cite{Ferrel,Anderson} or sample boundaries\cite{Schrieffer}, or to
accidental cancellations of the Fermi-contact  and core polarization
 contributions due
to peculiarities of the electronic structure. \cite{MacLaughlin}
\begin{figure}[ptb]
\rotatebox {270}{\includegraphics [width=2.3in]{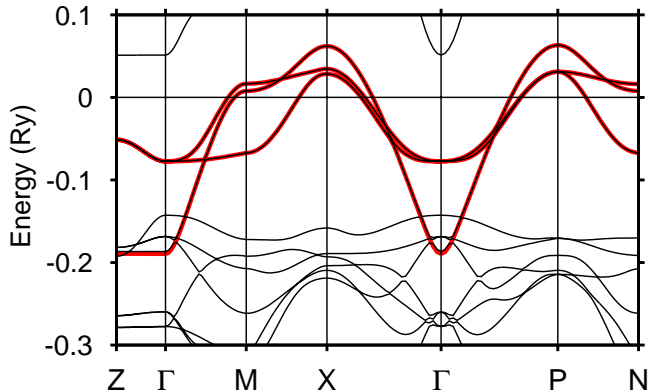}}
\caption{ (color online) LDA (NMTO) band structure (thin line) and pure $t_{2g}$ bands (thick line), obtained
by integrating out all other states (see Table \ref{tabhops}). The Fermi
level is set to zero. }%
\label{nmtobnds}%
\end{figure}

It is at this point clear that first principles calculations of the NMR Knight
shift and the relaxation rate are highly desirable, in order to gain better
understanding of the physics of the NMR in this compound. Similarly, to access
the applicability of alternative interpretations it is important to understand
better spin-orbit effects (note that selection between different $p$-wave
states is entirely controlled by spin-orbit coupling; see Refs.~\onlinecite{Mac,Ng2000}). 
In this paper we present full-potential highly accurate
calculations of the spin-orbit effects in Sr$_{2}$RuO$_{4}$ within the local
density approximation (LDA) with and without external magnetic field, as well
as first principles calculations  of the
NMR Knight shifts and the NMR relaxation rates, and analyze the possible
ramifications. In particular, we analyze the possibility of an accidental
cancellations due to electronic structure peculiarities, the possible
dominance of the van-Vleck term, and the effects of spin-orbit coupling and
correlations beyond LDA.

\section{Electronic structure}

Sr$_{2}$RuO$_{4}$ has the perovskite crystal structure of the cuprate
superconductor La$_{2}$CuO$_{4}$ (space group I4/mmm). \cite{Neumeier94} The
primitive cell is tetragonal body centered (lattice constants $a=b=3.87$%
\AA \ and $c=12.74$\AA ); Ru is located at (0,0,0) and Sr at (0,0,0.353),
while the two non-equivalent oxygens are O1 at (1/2,0,0) and O2 at
(0,0,0.1615) respectively.

\begin{figure}[ptb]
\rotatebox {0}{\includegraphics [width=3.25in]{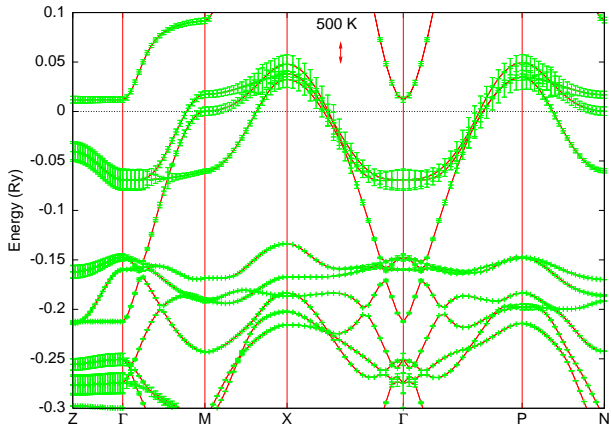}}\caption{
(color online) LDA (LAPW)
band structure with spin-orbit effects. The lines represent the actual band structure
(spin-orbit effects are small on this scale; the ``error bars'' indicate the size
of the spin-orbit induced shift of each state. In the central panel we show the length
of an ``error bar'' corresponding to a shift of 500 K ($\approx$ 43 meV).
The horizontal line is the Fermi energy.
}%
\label{sobnds}
\end{figure}

The electronic structure of Sr$_{2}$RuO$_{4}$ has been already studied by
several authors \cite{to95,ds95,im97II} by means of density functional theory
(DFT) in the local density approximation (LDA) or in the generalized gradient
approximation (GGA). In the present work we adopt different LDA approaches:
the Linear Muffin Tin Orbitals (LMTO) method in the atomic spheres
approximation (Stuttgart LMTO47 code),\cite{lmto} the N-th order Muffin Tin
Orbitals (NMTO) \cite{nmto} method and the NMTO-based downfolding approach
(Stuttgart NMTO47 code), which allows to calculate hopping integrals and
Wannier functions, and the linear augmented plane wave (LAPW) method (WIEN2k
code).\cite{WIEN2k} Our results for the band structure (Fig.~\ref{nmtobnds}
and Fig.~\ref{sobnds}) are consistent with previous calculations. The NMTO
bands (Fig.~\ref{nmtobnds}) are obtained from the self-consistent LMTO potential,
following a standard procedure. \cite{nmto}

\begin{table}[ptb]
\caption[{[tabhop}]{Hopping integrals $t^{lmn}$ in mRy. The connecting vector
is ${\mathbf{T}}=la\hat{x}+ma\hat{y}+nc\hat{z}$, where $a$ and $c$ are the
lattice constants; the hopping integrals are tabulated up to the fourth neighbors.
}%
\label{tabhops}
\begin {ruledtabular}
\begin {tabular}
{c|cccccccc} &$\epsilon _0$ & $t^{100}$ & $t^{010}$ & $t^{110}$ & $t^{\frac{1}{2}\frac{1}{2}\frac{1}{2}}$
&$t^{020}$&$t^{200}$\\
yz-yz &-26.80 & -3.03 & -22.59 & 0.94 & -1.18 &2.88 &0.06\\
xz-xz &-26.80 & -22.59 & -3.03 & 0.94 & -1.18 &0.06 &2.88\\
xy-xy &-25.21 & -27.73 & -27.73 & -8.18 & 0.09 &0.44& 0.44\\
yz-xz & 0 & 0 & 0 & 0.46 & -0.76 &0&0\\
yz-xy & 0 & 0 & 0 & 0 & 0.39 &0&0\\
xz-xy & 0 & 0 & 0 & 0 & 0.39 &0&0\\
\end {tabular}
\end {ruledtabular}
\end{table}

\begin{figure}[ptb]
\rotatebox {270}{\includegraphics [width=2.3in]{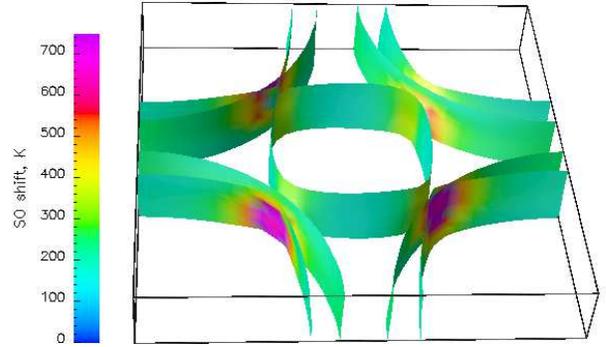}}\caption{
(color online) LAPW Fermi
surface, painted according to the  spin-orbit splitting at each point (see the color bar
for the scale in the temperature units).
}%
\label{soFS}%
\end{figure}

The tetravalent Ru has 4 electrons in the $d$ shell. The cubic crystal field
splits the $d$ levels into 3-fold degenerate $t_{2g}$ and 2-fold degenerate
$e_{g}$ states; the bands at the Fermi level are thus $t_{2g}$ bands 2/3
filled. DFT calculations show that these three t$_{2g}$ bands can be divided
into a wider $xy$ band, almost two dimensional, and two narrower almost
one-dimensional $xz$ and $yz$ bands. In Fig.~\ref{nmtobnds} we show the band
structure obtained by using the NMTO method. The t$_{2g}$ bands (thick line)
were obtained by integrating out all the degrees of freedom except for the
t$_{2g}$; by means of this first-principles downfolding procedure \cite{nmto}
we could construct Wannier functions and a real space Hamiltonian for these
bands\cite{njp3d1}. The corresponding hopping integrals are tabulated in Table \ref{tabhops}
up to the fourth nearest neighbors; farther hoppings are tiny and can be
neglected. The tables show that the inter-orbital hybridization is very small;
thus the bare band dispersion can be written as
\begin{align}
\label{H}\epsilon ({{\mathbf{k}}})  &  =\epsilon_{0}+2t^{100}\cos k_{x}%
+2t^{010}\cos k_{y}+4t^{110}\cos k_{x}\cos k_{y}\nonumber\\
&  +8t^{\frac{1}{2}\frac{1}{2}\frac{1}{2}}\cos\frac{k_{x}}{2}\cos\frac{k_{y}%
}{2}\cos\frac{k_{z}}{2}\nonumber\\
&  +2t^{200}\cos2k_{x}+2t^{020}\cos2k_{y}
\end{align}
for each band. 
These bare bands are slightly modified by the hybridization terms that 
take form

$$
H_{xz,yz}(\mathbf{k})=-4t^{110} \sin k_x\sin k_y-8t^{\frac{1}{2} \frac{1}{2} \frac{1}{2}}\sin\frac{k_x}{2}\sin\frac{k_y}{2}\cos \frac{k_z}{2}, 
$$

$$
H_{xz,xy}(\mathbf{k})=-8t^{\frac{1}{2} \frac{1}{2} \frac{1}{2}}\cos \frac{k_x}{2} \sin \frac{k_y}{2} \sin \frac{k_z}{2}, 
$$

$$
H_{yz,xy}(\mathbf{k})=-8t^{\frac{1}{2} \frac{1}{2} \frac{1}{2}} \sin \frac{k_x}{2} \cos \frac{k_y}{2} \sin \frac{k_z}{2},
$$
where $t^{lmn}$ are the hopping integrals tabulated in Table \ref{tabhops}.
This Hamiltonian
yields a Fermi surface consisting  of an electron-like sheet ($\gamma$), which is 
neraly cylindrical in shape, and four crossing planes, reconnected into a hole-like
($\alpha$) and an electron-like ($\beta$) tetragonal prisms by the weak
hybridization $xz-yz$. The Fermi surface obtained with the NMTO method is in
good agreement with previously reported results based on LAPW (Ref.~\onlinecite {im97II}). 
The LAPW Fermi surface is also displayed in 
Fig.~\ref{soFS}; qualitatively it is very similar to the results obtained in 
Ref.~\onlinecite {im97II}. Differently from previous calculations here we include
spin-orbit effects; these results will be discussed in Session V.

Since the bands at the Fermi level have mostly Ru $t_{2g}$ character with some
admixture of O $p$, it is likely that NMR is also dominated by Ru $d$ and O
$p$ contributions. In addition the polarization of the core electrons due to
the $d$ shell will be sizable for Ru. NMR results will be discussed in session IV.

\section{Stoner factor and spin-orbit interaction}

In the the spin density functional theory, the static spin susceptibility is
formally exact, and can be calculated as
\begin{equation}
\chi_{s}=dM_s(H)/dH, \label{chis}%
\end{equation}
where $M_s(H)$ is the spin magnetic moment induced by an external field $H.$
This is related to the \textit{non-interacting }spin susceptibility, which is
simply the density of states (DOS) in appropriate units, $\chi_{0}=\mu_{B}%
^{2}N(0),$ where $\mu_{B}$ is the Bohr magneton. The renormalization is
quantified in terms of the Stoner factor, $I:$
\begin{equation}
\chi_{s}=\chi_{0}/(1-I\chi_{0}/2). \label{RPA-M}%
\end{equation}
(in the literature one can find also an alternative definition of the Stoner
factor, per two spin instead of one, which differs from ours by the factor 
1/2 in the denominator).
One can also calculate the orbital part of the magnetic
susceptibility as
\begin{equation}
\chi_{L}=dM_{L}(H)/dH,
\end{equation}
as long as the spin-orbit interaction is included in the calculations.

\begin{figure}[ptb]
\rotatebox {0}{\includegraphics [width=3.5in]{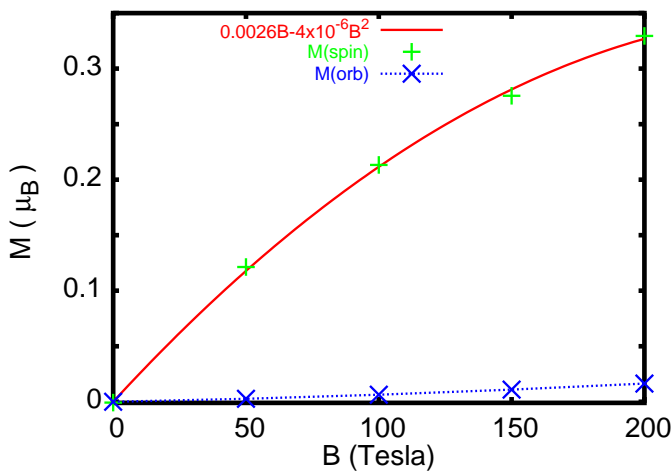}}
\caption{(color online) Orbital and spin magnetization (LAPW).}%
\label{chi-SO}%
\end{figure}

The most straightforward and the most accurate way to compute $I$ is by
performing self-consistent LDA calculations in an external field. For this
purpose we have selected the full-potential linear augmented plane wave method
as implemented in the WIEN2k+lo package.\cite{WIEN2k} This package allows for
self-consistent calculations in an external field, including interaction of
the field with the orbital moment. Two sets of calculation were performed, for
a field along $z$ or $x$ axes. Particular care has be taken to provide high
level of convergency: up to 729 inequivalent k-points (17x17x17 mesh), RKMAX
up to 9, states up to 3.5 Ry included in the second-variational
diagonalization of the spin-orbit Hamiltonian. Local orbitals were included in
the basis for both Ru and O, and a relativistic local $p$-orbital was added
for Ru, to ensure the convergence of the spin-orbit calculations. The results
are shown in Fig.~\ref{chi-SO} and Table \ref{tabchi}.
The following observations are in place: (1)
the spin susceptibility is anisotropic, with $z$ being the \textit{hard }axis, (2)
the orbital susceptibility is nearly isotropic and (3) there is a metamagnetic transition
at a field between 75 and 100 T to a state with a magnetic moment that
extrapolates to $\approx $0.1 $\mu_{B}/$Ru in zero field. While this metamagnetism
is obviously inobservable, it is probably related to the metamagnetism
observed in some Sr$_{2}$RuO$_{4}$-based alloys. In this paper we will not
discuss this metamagnetism and will concentrate on the susceptibility. The
low-energy slopes of $M(T)$ curves yield the results (in 10$^{-4}$ esu/mol)
listed in Table \ref{tabchi}. 
\begin{table}[tbp]
\caption[tabchi]{Static susceptibility in $10^{-4}\;$ emu/mol,
                         as obtained from LAPW band structure.}
\label{tabchi}%
\begin{ruledtabular}%
\begin{tabular}{l|llll}
& $\chi _{s},$ calc & $\chi _{L},$ calc & $\chi ,$
exp (Ref.~\onlinecite{Maeno97})  \\
$z$ & 7.26 & .48 & 9.8   \\
$xy$ & 7.67 & .43 & 8.8
\end{tabular}%
\end{ruledtabular}
\end{table}

We see that the calculations reproduce well the overall scale of the
susceptibility
but predict an opposite anisotropy compared to the
experiment, albeit of the same scale (6\% $vs.$ -11\%). Note that similar
problem was encountered by Ng and Sigrist in their model
calculation.\cite{Ng2000} They ascribed the discrepancy to anisotropy of the
orbital
 susceptibility, which, they pointed out should be larger in the
$z$ direction because of the possibility to form an $xz+iyz$ state out 
of degenerate $xy$ and $yz$ orbitals. Indeed, we find such an anisotropy
in our calculated orbital susceptibility, but it is way too small.
Yet we believe that Ng and Sigrist's conjecture was right and the 
reason we do not see this effect is due to the fact that 
the density functional calculations routinely underestimate orbital
effects (overestimate orbital moment quenching). 
In the experimental literature one can find variety
of estimates for the ratio $\chi_{orb}/\chi_{tot}$, from 10\%
\cite{Mac} to 17\% \cite{Ishida97}, while our result is 6\%.
Increasing our orbital susceptibility by a factor of two or three,
however, still does not reproduce the experimental anisotropy.
This strongly suggests, in accord with Ref. \onlinecite{Ng2000},
that the separation of the orbital and spin parts in the 
experiment was inaccurate and in reality the orbital
susceptibility is larger than usually quoted.
Indeed, the separation of the orbital and spin
susceptibilities in the experiment is based on the assumption
that the orbital part is totally temperature independent,
which is not true for metallic systems where a non-negligible
orbital susceptibility comes from the states at the Fermi level
(see, $e.g$, Ref. \onlinecite{Ziman}).
Note that the anisotropy of $\chi_{orb}$ extracted in
this way is 0.40$\times 10^{-4}$ emu/mol\cite{Ishida01},
2.5 times smaller than the observed anisotropy of the total
susceptibility\cite{Maeno97}.
We will return to this
issue again later when discussing the NMR experiments.

The calculation reported in this section yield a DOS of 50 states/Ry/cell. If
interpreted in terms of Eq.~(\ref{RPA-M}), this results in the Stoner factor
$I=0.46$ eV, and the renormalization coefficient $R=\chi_{s}/\chi_{0}=6.2.$ In
the next Section we will be using LMTO calculation which give slightly smaller
DOS, namely $45.6$ states/Ry/cell. We will therefore use such $R$ that
renormalizes the \textit{LMTO DOS }to the accurate LDA $\chi_{s},$ that is,
$R=6.8$. We point out that this value is about twice larger than the Stoner
enhancing factor we obtained directly from self-consistent
LMTO calculations in a magnetic field. This is not surprising; in our experience\cite{zrzn2}
allowing non-spherical variations of the spin density practically always results
in more magnetic solutions.

\section{NMR Knight shifts}

Next we calculate the NMR Knight shifts. 
These shifts originate from the hyperfine interaction $-\gamma_n {\bf I}\cdot {\bf H}$,
between the nuclear magnetic moment $-\gamma {\bf I}$ and the 
hyperfine field ${\bf {H}}$ produced at the nucleus site by the electrons.
The hyperfine field operator ${\bf {H}}$ is the sum of a core (${\bf H}^{core}$), a contact (${\bf H}^c$),
a dipole-dipole (${\bf H}^d$) and an orbital (${\bf H}^o$) term.
In order to calculate the NMR shifts we used two different approaches.

The first is described in Ref.~\onlinecite {nmrlmto} and based on the LMTO-ASA
method. Here we just summarize the most important steps. The bare Knight
shifts are calculated from the LDA band structure by using linear response
theory and without introducing any external magnetic field. 
\begin{table*}[tbp]\caption[tab1]{Knight shift, $K_\protect\protect\alpha $ in \%. The label $%
\protect\alpha =$x,y,z indicates the direction of the external magnetic
field.The first panel presents the results obtained with the LMTO method in the scalar relativistic approximation;
the Stoner enhancement factor $R=6.8$ is used for the renormalized values.
The second and the third panels are the results obtained with LAPW without and with spin-orbit (SO) interaction;
the Stoner factor is already included by construction.
The last panel lists available experimental data. For O1 the shifts for the (1/2,0,0) site are presented. }\label{tab1}
\begin{ruledtabular}\begin{tabular}{c|ccccccccccccccccc}
&\multicolumn{1}{c}{ dipole} &\multicolumn{1}{c}{contact}&orbital&
\multicolumn{1}{c}{ core} & \multicolumn{1}{c}{ Tot (xyz) (ren).}\\
&   x  \;\;\;\;\;\; y \;\;\;\;\;\; z &  & & & x \;\;\;\; \;\;y\;\;\;\;\;\; z\\
Ru    &  \phantom{-}0.007  \; \phantom{-}0.007  \;                -0.014     & 0.006   &0&-0.727   &-1.366 \; -1.366 \; -1.508 \\
{O1} &  \phantom{-}0.020  \;   -0.043 \;  \phantom{-}0.023       & 0.003   &0& \phantom{-}0.027   &\phantom{-}0.210 \; -0.218 \;\phantom{-} 0.231 
\\{O2} & \phantom{-}0.001  \; \phantom{-}0.001  \; -0.002       & 0.004  &0 & \phantom{-}0.002   & \phantom{-}0.038 \; \phantom{-}0.038 \;  \phantom{-}0.018           
\\{Sr} & \phantom{-}0.000 \;  \phantom{-}0.000 \; -0.001   & 0.059 &0   & -0.007 &\phantom{-}0.390 \;\phantom{-} 0.390 \;\phantom{-} 0.382           \\
\hline\hline
& dipole  &&orbital & {\ contact}+ {core} & Tot \\    
&   x  \;\;\;\;\;\; y \;\;\;\;\;\; z & &&  & x \;\;\;\; \;\;y\;\;\;\;\;\; z\\
Ru &   \phantom{-}0.238    \; \phantom{-}0.238   \; -0.475& &  0                           &     -1.437     &-1.200 \; -1.200 \; -1.912\\
O1 &   \phantom{-}0.216      \; -0.257 \;      \phantom{-}0.041   &&  0                        &    0.008          &    \phantom{-}0.224 \; -0.249 \; \phantom{-}0.048\\
O2 &   \phantom{-}0.021      \;  \phantom{-}0.021     \;   -0.042 & & 0                            &   0.015           & \phantom{-}0.036 \; \phantom{-}0.036 \; -0.027
\\Sr  &   \phantom{-}0.002    \;  \phantom{-}0.002   \;  -0.003 & &  0                    &    0.113         &\phantom{-}0.114 \; \phantom{-}0.114 \; \phantom{-}0.110\\
 \hline\hline
 & dipole  && orbital & {\ contact}+ {core} & Tot &  \\ 
 &   x  \;\;\;\;\;\; y \;\;\;\;\;\; z & &   x  \;\;\;\;\;\; y \;\;\;\;\;\; z&& x \;\;\;\; \;\;y\;\;\;\;\;\; z\\
 Ru &  \phantom{-}0.232    \; \phantom{-}0.232  \;  -0.463  && -1.146 \; -1.146  \;  -1.078   &    -1.436    &-2.350 \; -2.350 \; -2.977     \\  
 O1 & \phantom{-}0.211       \;  -0.254 \;    \phantom{-}0.043  &&   -0.104 \;  -0.067  \; -0.125   &       \phantom{-}0.006   &\phantom{-}0.113 \; -0.315 \; -0.076  \\  
 O2 &  \phantom{-}0.022   \; \phantom{-}0.022 \;  -0.043  &&     -0.088 \; -0.088 \; -0.117    &    \phantom{-}0.014   & -0.052 \; -0.052 \; -0.060   \\  
 Sr  &  \phantom{-}0.001    \; \phantom{-}0.001 \;  -0.003 &&      -0.138 \; -0.138  \; -0.134   &     \phantom{-}0.112   & -0.024  \; -0.024 \; -0.025  \\
\hline\hline
&Expt.\footnote{For Ru Ref.~\onlinecite{Ishida97}; for O1 ($T=4.2 K$) and O2 Ref.~\onlinecite{Imai98}.}
&Expt.\footnote{For Ru Ref.~\onlinecite{Ishida01,IshidaPC}. 
For O1 Ref.\onlinecite{Ishida98} ($T=2K$).}
&Expt. \footnote{For O1 and O2 Ref. \onlinecite{Mukuda98} (4.2 K).
}\\
 &   x  \;\;\;\;\;\; y \;\;\;\;\;\; z & x  \;\;\;\;\;\; y \;\;\;\;\;\; z& x \;\;\;\; \;\;y\;\;\;\;\;\; z\
\\{Ru} &  -2.75 \; -2.75 \; -3.44 & -2.70 \; -2.70 \; -3.32\\ 
{O1} &  \phantom{-}0.50 \; -0.15 \; \phantom{-}0.30     & \phantom{-}0.48 \; \phantom{-}0.18 \; \phantom{-0.00}&  \phantom{-}0.40 \; -0.15 \; \phantom{-}0.28&\\
{O2} &   \phantom{-}0.08 \; \phantom{-}0.08 \; 0.04 & &   \phantom{-}0.04 \; \phantom{-}0.04 \; \phantom{-}0.03 \\
{Sr} &   &&&    
\end{tabular}
\end{ruledtabular}\end{table*}
This means, e.g. for the spin terms,
$K \equiv 2 \mu_B \mathrm{Tr} \langle \uparrow | N {\bf {H}}|\uparrow \rangle$,
where we introduced the  density-of-states operator 
$\langle {\bf k} s | N |\mathbf{k}^\prime s^\prime \rangle\equiv \delta_{s s^\prime}
\delta{\mathbf{k}\mathbf{k}^\prime}\delta(\epsilon_{\mathbf{ k}}-E_F)$; here $s$ and $s^\prime$ are
spin indices, $\epsilon_{\bf k}$ the band energy and $E_F$  the Fermi level.
This approach
allows to clearly distinguish the contributions of the different orbitals and
of the different interactions and thus makes the physical interpretation
easier. The Stoner factor and the core shifts are obtained
separately, i.e. from a self-consistent LSDA calculation with an external
magnetic field, $B_{ext}$. The latter produces a core  polarization $m({\bf{r}})$;  
the core shift is the ratio
$B_{core}/B_{ext}$ between the (core) contact hyperfine field, $B_{core}=\mu_{B}%
(8\pi/3)m(0)$, and the external magnetic field. 
About 450 \textbf{k} points in
the irreducible Brillouin zone were needed in order to obtain well converged
results. We tested
several different LMTO set up \cite{setup}, with  consistent results.
The different contribution of conduction electrons to the Knight shift are
displayed in Table \ref{tab1}.

The second approach is the one implemented in the LAPW Wien2k package.
\cite{nmrwithwien} \emph{All} the contributions to the Knight were obtained
from the ratio between the hyperfine field $B$ and the external magnetic field $B_{ext}$
from which $B$ originates. Clearly, all the Knight shifts
calculated in this way are already Stoner renormalized. The Knight shifts were
obtained from the highly accurate LAPW band structure, and the effects of
spin-orbit could be accurately analyzed. The results are again displayed in
Table \ref{tab1}.

Let's examine first the results obtained for the Ru isotropic Knight shift,
$K_{iso}\equiv(K_{x}+K_{y}+K_{z})/3$ . We found that the $s$-electron
contribution from the conduction bands to $K_{iso}$ is negligible; this is
understandable because $N_{s}$, the Ru-$s$ projected density of states (see
Table \ref{tabdos}) is negligible at the Fermi level ($N_{s}/N_{t_{2g}}%
\approx 2\times 10^{-4}$).
The main contribution is the core polarization (the dipole-dipole term is
sizeable, but purely anisotropic).

At this point, it is interesting to compare the core-polarization Knight
shifts in LMTO and LAPW (no spin-orbit), Table \ref{tab1}. As already
explained, in both LMTO and LAPW the core polarization shifts were obtained from self-consistent
electronic structure calculations in an external magnetic field, and are
therefore renormalized by the Stoner factor. Still, the LMTO yields a core
shift a factor 1/2 too small; this is consistent with our results on the
Stoner enhancing factor $R$, which turned out to be a factor 2 smaller in LMTO
than in LAPW. For this reason, in calculating the Stoner renormalized shifts for the
LMTO (the rightmost column of Table \ref{tab1}) we took into account this
discrepancy and multiplied the core term by this missing factor 2. This is is
an important technical point and we would like to reiterate it: In LAPW
calculations all calculated contribution are by construction already 
Stoner-renormalized. In LMTO, the core polarization is renormalized by a
factor inherent to the ASA-LMTO calculation, which we found to be 
underestimated in ASA by 50\%, compared to our full-potential result.
 Spherical approximations for the crystal potential
commonly underestimate the tendency to magnetism compared to the full-potential
calculations in materials near a quantum critical point (cf., e.g., Ref.~\onlinecite{zrzn2}). 
The other contributions calculated in LMTO are found via linear
response formulas and do not include $any$ renormalization. Therefore, the
final result in the rightmost column of the Table \ref{tab1} includes a
renormalization by a factor of two for the LMTO core polarization, of 6.8 for
all other LMTO contributions, and no renormalization for the LAPW results.

We notice that the LAPW and the Stoner renormalized LMTO results for $K_{iso}$
(spin part) agree rather well; $K_{iso}\equiv RK_{iso}^{0}=-1.41$ in LMTO
and $-1.44$ in the non-relativistic LAPW (spin-orbit hardly affects this
term)\textbf{.} The small discrepancies can be attributed to small differences
in the band structure and in the calculation procedure. \cite{nmrwithwien}

\begin{table*}[ptb]
\caption[{[Table }]{ Projected density of states (states/Ry/atom),
obtained
with the LMTO method (set up with three empty spheres \cite{setup})
and comparison with LAPW.
In the last two columns $|\phi_{s}(0)|^{2}$ and
$\langle{1}/{r^{3}}\rangle_{l}\equiv\int drr^{2}|\phi_{l}(r)|(1/r^{3})$ 
(a.u.) are displayed, where $\phi_{l}(r)$ is the radial solution of
Schr\"odinger equation for the free atom and $l$ its angular quantum
number. }%
\label{tabdos}%
\begin{ruledtabular}
\begin{tabular}{c|ccccccccccccc}
 $K_c$& { N$_d$} & N$_{xy}$ & N$_{xz}$&N$_{yz}$& N$_s$  & 
 $\quad |\phi_s(0)|^2  \quad$ & $\quad \langle 1/ r^3\rangle_l\quad$\\
Ru (LMTO)  & 27.67 &10.33 &8.65 & 8.65   & 0.005& 3.46& 5.38 ($l=2$)\\
Ru (LAPW) & 29.28  & 14.45& 7.40 & 7.40&  0.002\\
\hline
& {N$_p$} & N$_x$ & N$_y$ & N$_z$ & { N$_s$}   &
$\quad |\phi_s(0)|^2  \quad$ & $\quad \langle 1/ r^3 \rangle_l\quad$   \\
O$_1$x (LMTO)& 5.54 & 0.02  & 2.44  &3.06& 0.010 & 8.00 & 5.15 ($l=1$)\\
O$_1$x (LAPW)& 6.30&0.02&3.90&2.39&0.011\\
O$_1$y (LMTO)& 5.54 & 2.44  & 0.02  &3.06& 0.010 & 8.00 & 5.15 ($l=1$)\\
O$_1$y (LAPW)& 6.30&3.90&0.02&2.39&0.011\\
O$_2$z (LMTO)&  0.88 & 0.36  & 0.36  &0.16& 0.015 &8.00 & 5.15 ($l=1$)\\
O$_2$z (LAPW) & 0.94 & 0.42 &0.42 &0.08&0.015\\
\end{tabular}%
\end{ruledtabular}
\end{table*}

The dipole contribution to $K_{iso}$ is zero by symmetry. However, the orbital
contribution is not, and it appears to be comparable to the contact term (as
opposed to the magnetic susceptibility, where the orbital part is rather
small). While nonrelativistic LMTO calculation cannot be used to evaluate this
contribution, the WIEN2k code includes an option of relativistic calculations
in an applied magnetic field. Importantly, we found that it was absolutely
necessary to include not only spin-orbit interaction, but interaction of the
orbital moment with the external field as well. We found that after including
the orbital part, $K_{iso}=-2.56$. Regarding the oxygen Knight shift, we get
$K_{iso}=-0.09$ and $-0.05$ for the in-plane and apical oxygen, respectively.

Experimentally, $K_{iso}\approx -2.98$ was reported in Ref.~\onlinecite{Ishida97}, 
and a slightly smaller
 value $K_{iso}\approx -2.91$ in later papers by the same group
\cite {Ishida01}.
Our calculated
number of $-2.56$ is 12\% smaller than the later data, reminiscent of
 the 15\% underestimation of the total susceptibility (Table \ref{tabchi}).
\cite{notevv}
 While the calculated $K_{iso}$ for oxygens has the wrong sign, this
does not seem disturbing, since both the experimental and the calculated
numbers are very small and result from a substantial cancellation of various
anisotropic terms, and the absolute magnitude of the error is rather small.

This  agreement seems impressively good, but, just as with the magnetic
susceptibility, this is probably deceptive. As mentioned, Sr$_{2}%
$RuO$_{4}$ is close to a magnetic quantum critical point, and therefore
belongs to a growing class of materials where LDA overestimates the tendency
to magnetism (see, $e.g.,$ Ref.~\onlinecite{Baku} and references therein).
 It is generally believed that magnetic fluctuations play an
important part in linear response in such materials,
suppressing the magnetic susceptibility compared to the essentially
mean field LDA treatment~\cite{qcpnote}. Indirect evidence indicates that LDA does indeed
overestimates the propensity to magnetism; for instance, in LDA Sr$_{2}%
$RuO$_{4}$ is unstable against formation of a spin density wave at the nesting
vector (1/3,1/3,0)\cite{halilov}. It is hard to imagine that the same
fluctuations that stabilize the paramagnetic state as opposed to the spin
density wave ground state in LDA would not reduce $\chi,$ $K$ (and also
$1/T_{1}T).$ Also, calculations predict Sr$_{2}%
$RuO$_{4}$ to be ferromagnetic at the surface\cite{plum}, while experimentally
 it is not the case\cite{andrea}.
 We conclude that the good agreement of the
calculated $\chi$ and $K_{iso}$ is largely due to a cancellation of errors: we
\textit{overestimate} the spin susceptibility, but \textit{underestimate} the
orbital susceptibility, as also evidenced by our failure to reproduce the sign
of the anisotropy in $\chi.$\cite{La} Note that underestimation of the orbital
polarization, due to correlation effects, is very typical for d-shells in the
transition metals.

Let us now analyze the anisotropy of the Knight shift. Experimentally,
Ishida {\it et al.}  \cite{Ishida97} reported 
for Ru  $K_{aniso}\equiv(K_{z}%
-K_{x})/3=-0.23\%$ and later
$-0.20\%$ (Ref.~\onlinecite {Ishida01,IshidaPC}).
Within the spherical approximation
for the potential, the LMTO calculations (Table \ref{tab1}) give $K_{aniso}%
=-0.05$ (which is solely a dipole effect). On the other hand, a full potential
approach (LAPW) results in $K_{aniso}=-0.23$ in the dipole part, which is
reduced to $-0.21$ by the orbital effects. This excellent agreement 
reflects the fact that the lion share of the calculated anisotropy
comes from the dipole part. 
Note that a possible overestimation of the contact interaction 
 would not affect  $K_{aniso}.$

It is instructive to analyze why LMTO underestimates the dipole contribution
compared to the full-potential calculations, by as much as a factor of 4.5 (for
Ru), despite the fact that it is is very accurate for the contact term. To
this end, let us recall that for the dipole interaction, the contributions
from $xy$ and $yz/xz$ bands  have opposite signs and tend to cancel each
other. Keeping only diagonal terms, which dominate, we find
\begin{equation}
\! K_{aniso}^{dip}\equiv\!\frac{K_{z}^{dip}}{2}\!\approx -\frac{\mu_{B}^{2}\langle
r^{-3}\rangle_{2}}{7}\left[  2N_{xy}\!-\!(N_{yz}+N_{xz})\right] .
\end{equation}
Here $N_{xy}$, $N_{yz}$ and N$_{xz}$ are orbital-projected density of states
at the Fermi level (Table \ref{tabdos}) per atom and $\langle r^{-3}\rangle
_{l}\equiv\int dr|\phi_{l}(r)|^{2}/r^{3}$, where $\phi_{l}(r)$ is the radial
solution of the Schr\"odinger equation with angular momentum quantum number
$l$. We found in ASA-LMTO (see Table \ref{tabdos}) that
$N_{xy}\approx  N_{xz}=N_{yz}$, 
and therefore the dipole
contributions nearly cancel each other. On the other hand, this cancellation
is far less complete in LAPW (see Table \ref{tabdos}): in LMTO the ratio $N_{xy}/N_{xz}\approx 1.2$,
while in LAPW it is nearly 2. Similar discrepancy is observed in the 
$N_x/N_y/N_z$ ratios for oxygens. On the other hand, the total DOSs, $N_d$
and $N_p$ agree reasonably well. Although one can expect that the 
spherical approximation somewhat underestimates the DOS anisotropy,
the observed discrepancies seem to be too large.
A close inspection
reveals the secret: LMTO underestimates the electrostatic crystal field
splitting, that is, the shift between the on-site energies of the $d_{xy}$ 
and $d_{xz}$ orbitals, by approximately 4 mRy, which leads to a relative shift
of the $xy$ and $xz/yz$ band by $\pm 4$ mRy compared to the LAPW calculations.
This seems like a small effect, and it is, but because the Fermi level is situated
on a slope of the DOS, shifting these bands in the opposite direction leads to
a serious redistribution of the DOS at $E_F$  between them. 
The more accurate NMTO bands, with a better account for nonspherical effects,
 are closer to LAPW: we find $N_{xy}-N_{xz}=4.9$ states/Ry/atom.

It is more complicated to analyze $K_{aniso}$ for oxygen, because of bigger
experimental uncertainty, strong temperature dependence (on the O1 site),
 and somewhat larger
orbital contribution in the calculations. The in-plane anisotropy on
the oxygen bridging two Ru along $x,$ 
 defined as $K_{aniso}^{\parallel}%
\equiv(K_{y}-K_{x})/2,$ is $-0.32\%$ in the experiment, while in the
calculation  it is  $-$0.21, mostly dipole. The out of plane anisotropy,
$K_{aniso}^{\perp}\equiv\lbrack K_{z}-(K_{x}+K_{y})/2]/3$, is 0.04 in the
experiment and 0.01 in the calculations. Again, one ought to keep in mind the
overall small magnitude. Finally, for the apical oxygen $K_{aniso}$ is
essentially zero both in the experiment and theory. Note that in case of
oxygen, LMTO also underestimates the dipole contribution, by virtue of the
same argument (for O, $K_{z}\approx-\frac{2}{5}\mu_{B}^{2}[2N_{z}-(N_{x}%
+N_{y})]\langle r^{-3}\rangle_{1}).$ 

\noindent\begin{table*}[tbp]
\caption[tab2]{$1/T_{1}T$ in (K sec)$^{-1}$, results based on LMTO calculations. 
The Stoner factors $R^{1.7}\approx 26$ and $2^{1.7}\approx
3.25$ (core) are used for the renormalized values (see text).
The 4th and 5th column display the total relaxation rate obtained after
Stoner renormalization of, respectively, the spin terms and of both spin and orbital terms.}
\label{tab2}%
\begin{ruledtabular}%
\begin{tabular}{c|ccccccccccc}
& orbital & dipole & {contact} & {core} &  Tot (spin renorm) &Tot (all renorm)
& Expt.
\footnote{For
$^{101}$Ru Ref.~\onlinecite{Ishida00,Imai98} (T=4.2 K).
For $^{17}$O (planar and apical)
data are from Ref.~\onlinecite{Imai98}
(T=4.2K).
}
&
Expt.
\footnote{Ref.~\onlinecite{Mukuda98}, sample with $T_c=1.5K$ ($T=4.2$K).}
\\
$^{101}$ Ru &  0.18692   &0.02855    &0.00004  & 0.53340                    & 2.67 &7.3           &15                   & \\  
$^{\phantom{0}17}$O1 &  0.03291  &0.01392    &0.00007  & 0.00541     & 0.41 & 1.23     &0.8   & 1.1\\  
$^{\phantom{0}17}$O2 &  0.00085    &0.00028    &0.00011  & 0.00004   & 0.011 & 0.032           & 0.025   & 0.025\\    
$^{\phantom{0}87}$Sr &   0.00029    & 0.00009    &0.00254  & 0.00003  & 0.069 &0.076           & -           &\\            
\end{tabular}%
\end{ruledtabular}
\end{table*}

Finally, strontium appears to be the only site where  the Fermi contact term
dominates. Unfortunately, to the best of our knowledge at present there are no
experimental data available for Sr.

\section{NMR relaxation rates}

\bigskip

Let us now analyze the relaxation rates. We calculated $1/T_{1}$ with the
procedure described in Ref.~\onlinecite {nmrlmto} and based on the LMTO band
structure.
Within this approach, the relaxation rate is obtained from  Fermi's golden rule
and  linear response theory;  for a polycrystalline sample $1/T_1T$ can
be expressed as
\begin{equation}
\frac{1}{T_1T}=2\pi k_B \hbar \gamma_n^2 \left[ \mathrm{Tr} \frac{1}{3} 
|\mathbf{H}N|^2\right] \label{rel},
\end{equation}
where $\mathbf{H}$ is the hyperfine field operator previously introduced.
Since the cross-terms in Eq.~(\ref{rel}) vanish exactly for a polycrystal \cite{nmrlmto}, 
the relaxation rate is, apart from the core polarization contribution,  the sum of contact, 
a dipole-dipole and an orbital term
\begin{equation}
\!\frac{1}{T_1T}\!=\!
2\pi k_B \hbar \gamma_n^2 
 \mathrm{Tr} \frac{1}{3}
\left[
|\mathbf{H}^cN|^2
+
|\mathbf{H}^dN|^2
+
|\mathbf{H}^oN|^2
\right]. 
\end{equation}
We calculated these contributions using the LMTO basis set; the results are displayed in Table \ref{tab2}.
The core polarization term was obtained separately from the core Knight
shift and Korringa law.

Available to us LAPW WIEN2k package currently does not allow for calculations
of the relaxation rate. However, we can make use of the understanding gained on
the previous stage by comparing LMTO and LAPW results for the Knight shifts to
evaluate the reliability of the LMTO calculations. 

We obtained the bare relaxation rate (Table \ref{tab2}) and calculated the
Stoner enhancing factor separately. The computation of the Stoner enhancing
factor for the relaxation rate, $R_{1/T_{1}T}$, requires knowledge of the
\textbf{q} dependence of the spin susceptibility, according to the following
formula\cite{Shastry94}:
\begin{equation}
R_{1/T_{1}T}\approx\left\langle \frac{{\mathit{Im}}\chi_{0}({\mathbf{q)}}%
}{[1-I({\mathbf{q)}}{\mathit{Re}}\chi_{0}({\mathbf{q)]}}^{2}}\right\rangle
/\left\langle {\mathit{Im}}\chi_{0}({\mathbf{q)}}\right\rangle .\label{RTT1}%
\end{equation}
Using the Lindhart susceptibility we showed earlier\cite{nmrlmto} that in 3D,
$R_{1/T_{1}T}\approx  R^{5/3}$. However, Sr$_{2}$RuO$_{4}$ is rather 2D, in which
case the real part of the noninteracting susceptibility is constant, and,
assuming a ${\mathbf{q}}$-independent $I,$ we get simply $R_{1/T_{1}T}\approx
R^{2},$ as first pointed out by Shastry and Abraham. \cite{Shastry94} Actual
dependence of $I$ on $q$ in Sr$_{2}$RuO$_{4}$ is discussed in Ref.~\onlinecite{im97II}, 
where it was estimated as $I(q)=I/[1+b(q/G)^{2}],$ where
$G=\pi/a$, and $b$ is a numerical constant close to 0.08. Using this with
Eq.~(\ref{RTT1}) and taking into account the actual $q$ dependence of
${\mathit{Im}}\chi_{0}({\mathbf{q)}}$ we find $R_{1/T_{1}T}\approx R^{1.7}%
=26$. 
We use this enhancement factor $R_{1/T_1T}$ to renormalize the contact and the dipole-dipole 
contributions to the relaxation rates.
Similar to the Knight shift, as discussed in Section IV,
the calculated core polarization contribution to the relaxation rate is
Stoner-renormalized according to the LMTO Stoner factor, which by itself
is smaller than the renormalization factor in LAPW. Thus, the core-polarization part
was renormalized by a factor $2^{1.7}$.

The renormalization of the orbital terms is more controversial.
Even if we knew the renormalization factor for the spin shifts exactly,
there is no guarantee that the renormalization for the orbital shifts is the same. 
Indeed, as discussed above in connection
with the Knight shift, the orbital polarization is induced by both direct interaction of the orbital
currents with the external field, as well as, indirectly, $via$ the spin-orbit interaction with the induced
spin density. The former process is not a subject of Stoner renormalization, while the latter is. 
It is therefore likely that the renormalization of the total orbital term is  not as large as the 
Stoner renormalization of the spin terms.
In Tab. \ref{tab2}  
we show the total relaxation rates obtained by renormalizing the spin shifts only, and those obtained renormalizing 
(with the same enhancing factor $\approx  26$) the orbital shifts too. 
These numbers are the lower and upper limit for the LDA relaxation rates.
Judging from the LAPW Knight shift (with spin-orbit)
one may expect that the  spin-orbit contribution to the relaxation rate is larger for
for Ru than for O; correspondingly, the renormalization of the orbital terms is
 expected to be larger for Ru than for O.

Since there are so many uncertainties in the renormalization of the orbital terms,
first we discuss the relaxation rates obtained  after renormalization
of the spin (contact, core and dipole-dipole) terms only.
The total calculated $1/T_{1}T\approx  2.67$ (K sec)$^{-1}$, underestimating the
low temperature
$1/T_{1}T\approx15\;$(K sec)$^{-1}$  by a factor 5.
For O1 we find large orbital and dipole-dipole terms, and a sizable core
polarization contribution. We find $1/T_{1}T=0.41\;$ (K sec)$^{-1}$.
Experimentally, $1/T_{1}T$ was found to be $0.8\;$ (K sec)$^{-1}$ at 4.2 K;
now the calculated value is too small by a factor 2. For O2 the value $1/T_{1}%
T= 0.023\;$ (K sec)$^{-1}$ was measured. \cite{Imai98,Mukuda98} Our calculated
value $1/T_{1}T=0.011\;$ (K sec)$^{-1}$, again to small of a factor 2.

In order to understand better the source of this  discrepancy 
we examine the case of Ru in more detail.
The leading contribution to $^{101} 1/T_1T$ is the core term;
judging from the Knight shift calculations, we do not expect  the core
polarization to be particularly poorly described by LMTO.
The  next largest term is the orbital contribution.
It is given, for Ru, for example, by the following expression
\begin{eqnarray}
\nonumber
\left(\frac{1}{T_1T}\right)_{orb} &=&\frac{4}{3}
C\left(\mu_B^2  \langle r^{-3} \rangle_2\right)^2 [ N_{xz}^2+N_{yz}^2  
\\ &+& 4N_{xy}^2+2N_{xy}(N_{xz}+N_{yz})], \label{orbt}
\end{eqnarray}
where $C\equiv~(4\pi k_B / \hbar)(\gamma_n / \gamma_e)^2\approx   0.404  \times 10^{5}/ (K s)$ for Ru$^{101}$ . 
Note that the orbital contribution to the relaxation rate does not {\em require} spin-orbit coupling 
(it appears already in the nonrelativistic approximation), but the spin-orbit coupling can and 
does
contribute to the orbital polarization. Note that if the spin-orbit contribution
to the relaxation rate would be of the same relative size as the corresponding
contribution to the Knight shift (that is, of the order of the core contribution)
we would expect an {\it additional} spin-orbit induced relaxation rate of
about $\approx  1.75 $ (K sec)$^{-1}$. This would be the {\it first} contribution
missing in Table  \ref{tab2}.

Furthermore,  the  orbital term is also partially underestimated because
of the difference between the LMTO and LAPW DOSs, as already discussed in the previous session.
Using LMTO partial DOSs in the approximate formula Eq.~(\ref{orbt})  we find 
$(1/T_1T)_{orb}\approx   0.29$ (K sec)$^{-1}$,  in good agreement with the full calculations
(Tab.~\ref{tab2}); using instead LAPW DOSs 
(the exact comparison of LAPW and LMTO is of course not possible because  space 
is divided up differently in the two methods)  we find
$(1/T_1T)_{orb}\approx   0.4 $ (K sec)$^{-1}$, which indicates that the LMTO orbital term is understimated 
by a factor $\approx  1.4$.  Similar considerations apply to the dipole-dipole term,
described by
\begin{eqnarray}
\nonumber
\left( \frac{1}{T_1T} \right)_{dip}&=&\frac{2}{49} 
C\left(\mu_B^2\langle r^{-3} \rangle_2\right)^2 
[N_{xz}^2+N_{yz}^2\\ &+& 4N_{xy}^2+3 N_{xy}( N_{xz}+N_{yz}) ].
\end{eqnarray}
This correction (for both terms) is about  0.4  (K sec)$^{-1}$.

Both these corrections together raise the estimate for the relaxation rate on Ru
to $\sim $ 5 (K sec)$^{-1}$, still short of the experimental number of 7.3 (K sec)$^{-1}$.
We take it to be yet another manifestation that the orbial effects on Ru are
underestimated in LDA and that the the leading relaxation mechanism is orbital.

One should expect that correlation effects on O are rather moderate and then,
after correcting for the LAPW-LMTO DOS differences, we should get a reasonable
agreement with the experiment (spin-orbit effects are also weak for O). Indeed, after 
this correction we obtain $(1/T_1T)_{O1}\approx 0.65 $ (K sec)$^{-1}$ and
$(1/T_1T)_{O2}\approx 0.16 $ (K sec)$^{-1}$, reasonably close to the experiment.

Finally, we neglected the quadrupolar contribution to $1/T_{1}T$.
For $^{101}$Ru this term is small, but not negligible \cite{noteQ}; 
taking this term into account would therefore increase the agreement
with experimental measurements. Interestingly, the quadrupolar term 
could partially contribute to smear out the Hebel-Slichter peak. 
The quadrupolar contribution is however  negligible for O1 and $^{99}$ Ru;
measurements of the relaxation rate below $T_{c}$ for these two ions could
therefore shade further light on our understanding of NMR data and clarify
the absence of the Hebel-Slichter peak.

\section{Superconductivity}

Finally, we want to make a few comments on possible alternative interpretation
of the invariance of the Knight shift across the superconducting
transition. 
As we discussed in the introduction (see also discussions in Refs.~\onlinecite{new,Dan}) 
the interpretation in terms of the order parameter, rotating  in
the field of 0.02 T, less that one per cent of the superconducting gap, is
very unlikely. The interpretation of the temperature independence of the
Knight shift in the superconducting state of Sr$_{2}$RuO$_{4}$ in terms of 
a chiral $p$-state is not possible, at least not on the level of existing theory.
Given that this is the \textit{only} experiment
uniquely identifying the superconductivity in Sr$_{2}$RuO$_{4}$ as
chiral $p$-wave, this opens the door for other possibilities regarding the symmetry of
superconducting state, such as the chiral $d$-state proposed in Ref.~\onlinecite{new}, 
$\Delta\propto xz\pm iyz,$ or a planar $p$-wave state, as 
suggested in Ref.~\onlinecite{Dan}.

Given the rather inexplicable absence of the Knight shift decay
below $T_{c},$ it is tempting to bring up an analogy with such materials as V
and Hg. Two explanations of the same phenomenon in
these elemental metals have
been proposed.
The first one is the spin-flip scattering\cite{Ferrel},\cite{Anderson} by the
grain boundaries or point defects due to the spin-orbit coupling. Spin-orbit
indiced splitting of the conductivity bands is rather nonuniform over the
Fermi surface, reaching 200K at some places (Figs.~\ref{sobnds}, \ref{soFS}). Anderson
\cite{Anderson} has derived a formula describing this effect: the ratio of the
Knight shift at $T=0$ and at $T>T_{c}$ is $K_{\sup}/K_{norm}\approx
\gamma_{s.f.}/6\Delta,$ where $\gamma_{s.f.}$ is the rate of the spin-flip
events and $\Delta$ is the superconducting order parameter, and $\gamma
_{s.f.}\agt\Delta$. If $\gamma_{s.f.}$ gets noticeably larger than
$\Delta,$ the Knight shift changes rather little across $T_{c}.$ In principle,
there is nothing unimaginable in $\gamma_{s.f.}>\Delta,$ however, we should
remember that nonmagnetic impurities in Sr$_{2}$RuO$_{4}$ are pair breakers,
and samples with $T_{c}\approx 1.4$ K, and the experiments we strive to explain
were performed on such samples, have to satisfy the pair breaking condition
$\gamma_{tr}<\Delta,$ where $\gamma_{tr}$ is the transport relaxation rate.
Even given the sizeable spin-orbit interaction in Sr$_{2}$RuO$_{4},$ it is
totally unrealistic to assume spin-flip relaxation to be larger than momentum
relaxation, that is $\gamma_{s.f.}\ll\gamma_{tr}<\Delta$ should hold.

Another mechanism that can emulate a temperature independent Knight shift
was described by MacLaughlin\cite{MacLaughlin}: if the Fermi contact
and the core polarization contribution cancel each other, the remaining
orbital part, being an entirely charge effect, will stay unsupressed 
below $T_c$. This was supposed to be operative, for instance, in V.
There are two objections against applying the same scenario to Sr$_{2}$RuO$_{4}$.
First, in Ru we find that the Fermi contact interaction is negligible, but
the core polarization is responsible for about a half of the total Knight 
shift.
Even though it is probably overestimated, we can guarantee it is not negligible.
Moreover, for both O1 and Ru we find a sizeable dipole term (in fact, for O1 it is
the leading term). In Ref.~\onlinecite{MacLaughlin} this term was neglected, but
it also originates from the spins of electrons and should be sensitive to pairing.
So, on this (probably oversimplified, as discussed below) level neither of the
two known explanations of the invariance of the Knight shift below $T_c$ is
applicable.

While this simplistic picture\cite{MacLaughlin} does not appear to work for Sr$_{2}$RuO$_{4}$
it brings up an important point, that the standard theory of the Knight shift
in the superconducting state, developed for Fermi-contact interaction in one-band
superconductors has to be revised to be applicable to Sr$_{2}$RuO$_{4}$.
Indeed, while the orbital interaction does not dominate the Knight shift on Ru,
it constitutes at least a half of the total $K$, and probably as much as 2/3, if
we take into account that the core polarization is probably overestimated in LDA.
This alone raises our expectation for $K(T=0)$ from zero to about 2/3 of its 
normal value. On top of that we have a sizeable dipole term, which for the in-plane
field is opposite in sign to the core polarization. Most interestingly, the  temperature
dependence below  $T_c$ should be highly nontrivial even for a singlet pairing: 
as discussed, this term comes about through near-cancellation of the contributions
from the $xy$ and $xz+yz$ bands. For several reasons, these bands are expected
to react differently to magnetic field (and the actual measurements were performed
in a sizeable field): they have different Fermi velocities along the $c$-axis, and 
possibly rather different gap values. Therefore, the partial densities of states 
decay (in a singlet case) below $T_c$ at different rates, with a possibility of the net
dipole Knight shift to decrease, increase or (accidentally) stay constant! The 
exact answer requires a quantitative analysis that goes beyond the framework of 
this paper, but we want to emphasize that {\it the fact that the dipole contribution
is strong calls for an entirely new theory of the Knight shift below $T_c$}.
A similar consideration applies to the Knight shift on O.

\section{Conclusions}

We have calculated from first principles the magnetic susceptibilty, the
Knight shifts and the relaxation rates in Sr$_{2}$RuO$_{4}$. We adopted
different LDA techniques (LMTO and LAPW) and performed calculations with and
without explicit account for the spin-orbit interaction. The agreement with
available experimental data is good, in some cases surprisingly good.
The isotropic magnetic susceptibility, the isotropic and anisotropic
Knight shift for Ru are all reproduced better than within 15\%.
However, there are notable discrepancies,
of which most important is, probably, the incorrect anisotropy of magnetic
susceptibility. 
We argue that
this error is related to an enhancement of the orbital polarization due to
correlation effects beyond LDA.
Despite the good quantitative agreement with the experiment of the
isotropic suseptibility and isotropic Knight shift on Ru, we do think that we
witness a cancellation of two errors: On one hand, we overestimate the spin
susceptibility, which should be reduced by spin fluctuations missing in the
LDA, and on the other hand we underestimate the orbital polarization in the Ru
d shell. If this interpretation is correct, one has to conclude that orbital
effects are extremely important and probably dominate both the Knight shift and the NMR
relaxation on Ru.

We have considered the accepted mechanisms of the Knight shift invariance in
elemental metals and conclude that they cannot explain the same phenomenon 
in Sr$_{2}$RuO$_{4}$. In particular, we can exclude 
the spin-flip scattering by defects in Sr$_{2}$RuO$_{4}$.
On the other hand, orbital polarization of electrons at the Fermi level,
which was used earlier to explain a similar effect in vanadium, is
also the leading contribution in  Sr$_{2}$RuO$_{4}$. Although it 
does not entirely dominate the Knight shift, as in V, it does substantially
reduce the expected effect of singlet superconductivity on it.
Finally, sizeable contribution of the dipole-dipole interactions together
with the orbital-dependent order parameter
essentially renders the existing theory of the Knight shift suppression
by a singlet pairing incomplete and calls for its revision.
We hope that this conclusion will inspire further experimental and
theoretical studies.

\section{Acknowledgments}

We acknowledge fruitful discussions with D. Agterberg. We also wish to
thank P. Novak for consultations concerning the WIEN2k
implementation of the Knight shifts calculations. E.P. acknowledges
INFM-Iniziativa calcolo parallelo for support.


\begin{thebibliography}{99}
\bibitem {setup}The results presented were obtained for a set up with three
empty spheres ($R_{Sr}=3.28$, $R_{Ru}=2.44$, $R_{O1}=1.8$, $R_{O2}=2.06$,
$R_{E1}=1.92$, $R_{E2}=1.32$, $R_{E3}=1.26$) .
\bibitem {nmrwithwien}In Wien2k the Knight shift are presently calculated
retaining only the $l$-diagonal terms and only the contribution from the
muffin-tin spheres. In the case of Sr$_{2}$RuO$_{4}$, detailed inspection of
the LMTO NMR calculations, which take into account the off-diagonal terms,
suggests that the latter are small.
\bibitem {notevv} The fact that both experimental and calculated
susceptibilities and Knight shifts include several contribution
of different physical origins complicates the comparison with the
experiment. It is tempting to compare, for instance, our calculated spin
susceptibility of 7.5 $\times 10^{-4}$ with the published number
of $\chi_s \approx 7.6 \times 10^{-4}$ (Ref.\onlinecite{Ishida97})
and claim an excellent agreement. However, as mentioned before,
the standard procedure used in processing the
experiment neglects possible temperature dependence of the orbital
susceptibility, which is questionable in case of orbital contributions
from electrons at the Fermi level. Besides, it makes use of free Ru$^{4+}$
ion calculations of the orbital hyperfine field with a semiempirical
reduction coefficient. On the other hand, we calculated the orbital
contribution directly, but only from Ru 3d and O 2p electrons. Still,
we believe that other contributions are small and a comparison
of our calculations should be with the $total$ experimental 
susceptibilities and Knight shifts.
\bibitem{qcpnote}
 At early stages, SrRu$_2$O4 was believed to be close to a ferromagnetic 
instability \cite{Rice95,im97II}. Ferromagnetic quantum critical spin-fluctuations  were believed to be responsible
for triplet superconductivity \cite{Rice95,im97II}, and even for stabilizing the particular
chiral p-wave state \cite{Sigrist99}.
While it was later discovered \cite{Braden} tht Sr$_2$RuO$_4$ is actually closer to an AFM spin-density
wave instability (which makes stabilization of a p wave state rather puzzling), it is generally accepted that
critical fluctuations are very important for superconductivity in this compound.
\bibitem {noteQ} An estimate of the quadrupolar contribution to the relaxation
rate with respect to the spin and orbital terms is the ratio $(Q/\mu)^{2}$,
which is about 0.3 for $^{101}Ru$. This ratio is $\approx 0.01$ for $^{17}$O.
\end{thebibliography}
\end{document}